\documentclass[preprint,showpacs,showkeys,preprintnumbers,amsmath,amssymb]{revtex4}
\usepackage{epsf}
\usepackage{color}
\usepackage{dcolumn}
\usepackage{bm}
\everymath{\displaystyle}
\begin{document}

\title{Rotating charged Black Holes in Einstein-Born-Infeld theories and their ADM
mass.}
\author{Diego Julio Cirilo Lombardo}
\address{Bogoliubov Laboratory of Theoretical Physics, JINR, 141980 Dubna,\\
Russia}
\date{\today}

\begin{abstract}
In this work, the solution of the Einstein equations for a slowly rotating
black hole with Born-Infeld charge is obtained. Geometrical properties and
horizons of this solution are analyzed. The conditions when the ADM mass (as
in the nonlinear static cases) and the ADM angular momentum of the system
have been modified by the non linear electromagnetic field of the black
hole, are considered.
\end{abstract}
\maketitle

\section{Introduction}

The four dimensional solutions with spherical symmetry of the Einstein
equations coupled to Born-Infeld fields have been well studied in the
literature[1]. In particular, the electromagnetic field of the Born-Infeld
monopole, in constrast to Maxwell counterpart, contributes to to the ADM
mass of the system (i.e. the four momentum of assymptotically flat
manifolds) . B. Hoffmann was the first who studied such static solutions in
the context of the general relativity with the idea to obtain a consistent
particle-like model. Unfortunately, these static Einstein-Born-Infeld models
generate conical singularities at the origin [1, 9]. This type of
singularities cannot be removed as global monopoles or other non-localized
topological defects of the spacetime [6, 7]. More recently, the solutions in
Einstein-Born-Infeld (EBI) theory become significatives because in open
superstring theory [13, 14], loops calculations lead the BI\ Lagrangian with
$b=\frac{1}{2\pi \alpha ^{\prime }}$ , where $\alpha ^{\prime }$is the
inverse of the string tension and $b$ the Born Infeld parameter.

In this report, the solution for slowly rotating black hole with Born-Infeld
charge is obtained. This solution, assymptotically flat, presents non linear
terms that modifys assymptotically the mass and the angular momentum (ADM
values) of the spacetime. Families of solutions are obtained varying $b$. It
is well known that the mass and angular momentum in the Kerr-Newman model of
the spacetime appear as integration constants that correspond to the ADM \
values of this geometry. In our non linear model of rotating charged black
hole the ADM\ mass and angular momentum are not zero when we put these
integration constants (that are required in the Kerr Newman's model) equal
to zero. For particular values of $b$ there are solutions with $%
M^{2}<a^{2}+Q^{2}$ without naked singularities.

In the Einstein-Born-Infeld model of a rotating black hole, one expects to
find a metric with an assymptotic behaviour as the well known Kerr-Newman's
metric. There are few difficulties: the metric is non-diagonal (rotating
frame) and the energy-momentum tensor of Born-Infeld includes the invariant
pseudoscalar [2, 3] (the magnetic field appears because of the rotation of
the compact object). The used convention is the {\bf spatial} of Landau and
Lifshitz (1962), with signatures of the metric, Riemann and Einstein tensors
all positives (+++)[4, 5]

\section{Statement of the problem}

The theory starts with a Born-Infeld field interacting with gravity and is
described by the action
\[
I=\int d^{4}x\left( \sqrt{g}R-{\cal L}_{BI}\right)
\]
where:

\[
{\cal L}_{BI}=\sqrt{g}L_{BI}=\frac{b^{2}}{4\pi }\left\{ \sqrt{-g}-\sqrt{%
\left| \det (g_{\mu \nu }+b^{-1}F_{\mu \nu })\right| }\right\}
\]
that with the expansion of the determinants in four dimensions, is easy to
obtain the standard form of the Born-Infeld lagrangian:
\[
L_{BI}=\frac{b^{2}}{4\pi }\left\{ 1-\sqrt{1+\frac{1}{2}b^{-2}F_{\mu \nu
}F^{\mu \nu }-\frac{1}{16}b^{-4}\left( F_{\mu \nu }\widetilde{F}^{\mu \nu
}\right) ^{2}}\right\}
\]
In an orthonormal frame (tetrad) the above given lagrangian becomes:

\[
L_{BI}=\frac{b^{2}}{4\pi }\left( 1-\sqrt[2]{1-\frac{2S}{b^{2}}-\frac{P^{2}}{%
b^{4}}}\right)
\]
where we have defined the invariants of the electromagnetic tensor $F$ as
\[
S\equiv -\frac{1}{4}F_{ab}F^{ab}=L_{M}
\]
\[
P\equiv -\frac{1}{4}F_{ab}\widetilde{F}^{ab}
\]
with the conventions
\[
\widetilde{F}^{ab}=\frac{1}{2}\varepsilon ^{abcd}F_{cd}
\]
\[
a,b,c....\equiv Tetrad\ \ indexes
\]

We consider the following ans\"{a}tz for the line element ( as the Boyer and
Linquist interval [8, 9] for the Kerr-Newman's geometry) with the expected
assymptotic behaviour:
\begin{equation}
ds^{2}=-\frac{\Delta }{\rho ^{2}}\left[ dt-asin^{2}d\varphi \right] ^{2}+%
\frac{sin^{2}\theta }{\rho ^{2}}\left[ \left( r^{2}+a^{2}\right) d\varphi
-adt\right] ^{2}+\frac{\rho ^{2}}{\Delta }dr^{2}+\rho ^{2}d\theta ^{2}
\end{equation}
where the functions to be determined $\Delta $ and $\rho $, are in principle
depending on $r$ and $\theta $. To obtain the Einstein equations, the most
powerful is the Cartan's method [5, 8, 9]. This method applies differential
forms and is based on two fundamental geometric equations (structure
equations). In the orthonormal basis of 1-forms the line element is written:
\[
ds^{2}=-(\omega ^{0})^{2}+(\omega ^{1})^{2}+(\omega ^{2})^{2}+(\omega
^{3})^{2}
\]
where the association between coordinate and orthonormal frame is not
trivial in the case of axially rotating symmetry and requires solving an
equation system. Explicitly, the tetrad is:
\[
\omega ^{0}=\frac{\sqrt{\Delta }}{\rho }\left( dt-a.sin^{2}\theta d\varphi
\right)
\]
\[
\omega ^{1}=\frac{sin\theta }{\rho }\left[ \left( r^{2}+a^{2}\right)
d\varphi -adt\right]
\]
\[
\omega ^{2}=\frac{\rho }{\sqrt[2]{\Delta }}\ dr
\]
\[
\omega ^{3}=\rho \ d\theta
\]
.

For the electromagnetic tensor $F$, we propose a similar structure to the $F$
of the Boyer and Lindquist generalization for the Kerr-Newman problem [8,
9]:
\begin{eqnarray*}
F &=&F_{20}dr\wedge \lbrack dt-asin^{2}\theta d\varphi ]+F_{31}sin\vartheta
d\theta \wedge \lbrack \left( r^{2}+a^{2}\right) d\varphi -adt] \\
&=&F_{20}\omega ^{2}\wedge \omega ^{0}+F\omega ^{3}\wedge \omega ^{1}
\end{eqnarray*}
where $F_{20}$ and $F_{31}$ are to be determined. We can see that $F_{20}$\
and $F_{31}$\ are the only field components in the tetrad.

Next we find the energy momentum tensor components in the rotating system $%
\left( tetrad\right) $. We shall use the metric symmetrized expression of $%
T_{b}^{a}$:
\[
T_{\ \ b}^{a}=\delta _{\ \ b}^{a}{\cal L}_{BI}{\cal \ }-\frac{\partial {\cal %
L}_{BI}}{\partial S}F_{\ \ \ l}^{a}F_{\ \ b}^{l}-\frac{\partial {\cal L}_{BI}%
}{\partial P}F_{\ l}^{a}\widetilde{F}_{\ b}^{l}
\]
In our case, the energy-momentum tensor takes a diagonal form
\[
-T_{00}=T_{22}=\frac{b^{2}}{4\pi }\left( 1-u\right)
\]
\begin{equation}
T_{11}=T_{33}=\frac{b^{2}}{4\pi }\left( 1-u^{-1}\right)
\end{equation}
where:
\[
u\equiv \sqrt{\frac{(\overline{F}_{31})^{2}+1}{1-\left( \overline{F}%
_{02}\right) ^{2}}}\,\ \ \ \ \ ;\ \ \ \ \ \ \left( \overline{F}_{ab}\equiv
\frac{F_{ab}}{b}\right)
\]
As the geometrical symmetries of the Riemann tensor
\[
R_{\alpha \mu \nu \lambda }+R_{\alpha \nu \lambda \mu }+R_{\alpha \lambda
\mu \nu }=0
\]
and the well known Bianchi identities
\[
R_{\alpha \beta \mu \nu ;\lambda }+R_{\alpha \beta \nu \lambda ;\mu
}+R_{\alpha \beta \lambda \mu ;\nu }=0
\]
$\rho $\ inmediatelly can be determinated:
\[
\rho ^{2}=r^{2}+a^{2}\cos ^{2}\theta =\Sigma
\]
We can see without losing generality that the function $\rho $ is the same
as the $\rho $\ of Boyer and Linquist and does not depend on the axially
symmetric source considered. With the function $\rho $\ found, only $\Delta $%
\ left to be found.

From the Einstein equations with the components (2) of the energy-momentum
tensor of Born-Infeld in the tetrad [9, 12], we obtain the following
expression
\begin{equation}
2\left( 8\pi \right) (T_{11}+T_{22})=-\frac{2}{\rho ^{2}}+\frac{\partial
_{r}\partial _{r}\Delta _{\left( r,\theta \right) }}{\rho ^{2}}
\end{equation}
that must to be solved with the following boundary conditions:
\[
\lim_{r\rightarrow \infty }\Delta _{RBI}\rightarrow \Delta _{Kerr-Newman}
\]
that will give us an assymptotically flat solution, with the correct
Maxwellian behaviour at great distances of the non linear source of the
electromagnetic fields; and
\[
\Delta \left( r_{h}\right) =0\,\ \ \ \ \ \ \ with\,\ \ \ \ \left. \
R_{\alpha \beta \gamma \delta }R^{\alpha \beta \gamma \delta }\right|
_{r=r_{h}}\neq \infty
\]
It is the usual regularity condition for the horizon.

Notice that the expression (3) for the well known Kerr-Newman model takes
the simply form
\[
2=\partial _{r}\partial _{r}\Delta _{\left( r\right) }
\]
because the energy-momentum tensor in the Maxwell theory is traceless.
Notice also, from the expression given above, that in the Kerr-Newman model
the ADM mass $M_{ADM}$, the angular momentum $a_{KN}$ and the charge $Q$
must to be appear necessarily as integration constants. We will show, that
is not the case in the BI theory, and these parameters arises naturally .

The fields will be obtained from the dynamical (eulerian) equations in the
tetrad form. Let us solve these equations with the following boundary
condition: the fields assymptotically have the same behaviour as the
electromagnetic fields of the Kerr-Newman model.
\[
\nabla _{a}\left[ \frac{1}{b^{2}\text{{\sc R}}}F^{ab}+\left( \frac{P}{b^{4}%
\text{{\sc R}}}\right) \widetilde{F}^{ab}\right] =0
\]
with the Maxwellian asymptotical behaviour of the fields, explicitly
\[
\left. F^{20}\right| _{r\rightarrow \infty }\rightarrow -\frac{Q}{\rho ^{4}}%
\left( r^{2}-a^{2}\cos ^{2}\theta \right)
\]
\[
\left. F^{31}\right| _{r\rightarrow \infty }\rightarrow \frac{2Q}{\rho ^{4}}%
ar\cos \theta
\]

Then we obtain ($r_{o}=\sqrt{\frac{Q}{b}}$):
\[
\left( \overline{F}_{20}\right) ^{2}=\frac{r_{o}^{4}r^{2}}{\rho ^{8}\left(
r_{o}^{4}+r^{4}\right) }\left[ r^{2}\left( r^{2}-a^{2}\cos ^{2}\theta
\right) ^{2}-4a^{2}\cos ^{2}\theta r_{o}^{4}\right]
\]
\[
\left( \overline{F}_{31}\right) ^{2}=\frac{\left( r_{o}^{4}+r^{4}\right) }{%
r_{o}^{4}r^{2}}.\frac{\left[ 4.\cos ^{2}\theta r^{2}a^{2}r_{o}^{8}\left(
r^{2}-a^{2}\cos ^{2}\theta \right) ^{2}\right] }{\left[ r^{2}\left(
r^{2}-a^{2}\cos ^{2}\theta \right) ^{2}-4a^{2}\cos ^{2}\theta r_{o}^{4}%
\right] \rho ^{8}}
\]

Putting all the ingredients in the equation $\left( 3\right) $, we obtain
the following expression:
\[
4\rho ^{2}b^{2}-2\rho ^{2}b^{2}\left[ \sqrt[2]{\frac{\left(
r_{o}^{4}+r^{4}\right) }{\rho ^{4}r^{2}-4a^{2}\cos ^{2}\theta \left(
r_{o}^{4}+r^{4}\right) }}\frac{\left( r^{2}-a^{2}\cos ^{2}\theta \right) }{r}%
\right] -
\]
\[
-2\rho ^{2}b^{2}\left[ \sqrt[2]{\frac{\rho ^{4}r^{2}-4a^{2}\cos ^{2}\theta
\left( r_{o}^{4}+r^{4}\right) }{\left( r_{o}^{4}+r^{4}\right) }}\frac{r}{%
\left( r^{2}-a^{2}\cos ^{2}\theta \right) }\right] +2=\partial _{r}\partial
_{r}\Delta
\]

This expression, although exact, is not integrable by trascendental
functions as in the static cases. One must make an expansion in power series
for small (slowly rotating) $a/r$
\[
4\rho ^{2}b^{2}-2b^{2}\left\{ \sqrt[2]{r_{o}^{4}+r^{4}}\left[ 1+\frac{a^{2}}{%
r^{2}}\cos ^{2}\theta \left( 1+\frac{2r_{o}^{4}}{r^{4}}\right) -3\frac{a^{4}%
}{r^{4}}\cos ^{4}\theta \left[ 1+\frac{2r_{o}^{4}}{r^{4}}\left( 1+\frac{%
r_{o}^{4}}{r^{4}}\right) \right] \right] +\right.
\]
\begin{eqnarray*}
\left. \frac{r^{4}}{\sqrt[2]{r_{o}^{4}+r^{4}}}\left[ 1+\frac{a^{2}}{r^{2}}%
\cos ^{2}\theta \left( 1-\frac{2r_{o}^{4}}{r^{4}}\right) -\frac{a^{4}}{r^{4}}%
\cos ^{4}\theta \left[ 1+\frac{2r_{o}^{4}}{r^{4}}\left( 1-\frac{r_{o}^{4}}{%
r^{4}}\right) \right] \right] \right\} +2 &=&\partial _{r}\partial
_{r}\Delta \  \\
&&\left( \frac{a}{r}<<1\ ;\ r_{0}\lesssim r\right)
\end{eqnarray*}
This expansion does not affects the regularity condition of the horizon.
Looking at the last equation,one can see that $\Delta $ depends on the
radial coordinate $r$ and the angular coordinate $\theta $. The integrals
are calculated in the indefinite form and the values of the two constants $%
A\left( \theta \right) $ and $B\left( \theta \right) $ of the problem are
selected according to the assymptotical flat behaviour of $\Delta $ and the
metric. Is useful to see, previously, an intermediate computation
\begin{eqnarray*}
\partial _{r}\Delta &=&2r+\frac{2}{3}b^{2}\left\{ 2\left( r^{3}-r\sqrt{%
r^{4}+r_{0}^{4}}\right) +\sqrt{i}\ r_{0}^{3}\ F\left[ iArc\text{Sinh}\left[
\left( -1\right) ^{1/4}\frac{r}{r_{0}}\right] ,-1\right] \right\} + \\
&&+b^{2}\ a^{2}\cos ^{2}\theta \left[ r+\sqrt{r^{4}+r_{0}^{4}}\left( \frac{%
4r_{0}^{4}}{5r^{5}}-\frac{2}{5r}-\frac{54\ a^{2}\cos ^{2}\theta }{11r^{3}}-%
\frac{16\ r_{0}^{4}\ a^{2}\cos ^{2}\theta }{11r^{7}}-\frac{12\ r_{0}^{8}\ \
a^{2}\cos ^{2}\theta }{11r^{11}}\right) \right] \ - \\
&&-\left[ \frac{18\ r_{0\ }b^{2}\ a^{2}\cos ^{2}\theta }{5\sqrt{i}}\left( E%
\left[ iArc\text{Sinh}\left[ \left( -1\right) ^{1/4}\frac{r}{r_{0}}\right]
,-1\right] -F\left[ iArc\text{Sinh}\left[ \left( -1\right) ^{1/4}\frac{r}{%
r_{0}}\right] ,-1\right] \right) \right] - \\
&&-\frac{34\ \sqrt{i}_{\ }b^{2}\ a^{2}\cos ^{2}\theta }{11r_{0}}F\left[ iArc%
\text{Sinh}\left[ \left( -1\right) ^{1/4}\frac{r}{r_{0}}\right] ,-1\right]
+A\left( \theta \right)
\end{eqnarray*}
The obtained solution takes the following form:
\begin{eqnarray*}
\Delta \left( r,\theta \right) &=&r^{2}+P_{ST}+\left\{ 2a^{2}\frac{Q^{2}}{%
\left( r_{0}\right) ^{3}}\left\{ \cos ^{2}\theta \frac{9}{5}\sqrt{i}E\left[
\frac{\pi }{4},2\right] +1.525\sin ^{2}\theta \right\} -2M+\right. \\
&&+\left. 2\frac{Q^{2}}{\left( r_{0}\right) ^{5}}a^{4}\left( 2\sin
^{2}\theta -\sin ^{4}\theta \right) 2.8653\right\} r+a_{KN}^{2}+ \\
&&+2\left( \frac{a}{r_{0}}\right) ^{4}Q^{2}\left( 2\sin ^{2}\theta -\sin
^{4}\theta \right) 0.8576+ \\
&&+2a^{2}\frac{Q^{2}}{\left( r_{0}\right) ^{4}}\cos ^{2}\theta \left\{ r^{2}+%
\frac{4}{5}\sqrt{r_{0}^{4}+r^{4}}-\frac{1}{10}\left( \frac{r_{0}}{r}\right)
^{4}\sqrt{r_{0}^{4}+r^{4}}-\right. \\
&&-\left. rr_{0}\frac{9}{5}\sqrt{i}E\left[ \frac{1}{2}ArcCos\left[ i\left(
\frac{r}{r_{0}}\right) ^{2}\right] ,2\right] \right\} +2\left( \frac{a}{r_{0}%
}\right) ^{4}Q^{2}\cos ^{2}\theta \left\{ -2Arc\text{Sinh}\left( \frac{r^{2}%
}{r_{0}^{2}}\right) \right. + \\
&&+\left. \sqrt{1+\left( \frac{r}{r_{0}}\right) ^{4}}\left[ -\left( \frac{%
r_{0}}{r}\right) ^{2}\frac{163}{770}-\left( \frac{r_{0}}{r}\right) ^{6}\frac{%
313}{385}-\left( \frac{r_{0}}{r}\right) ^{10}\frac{21}{385}\right] \right\} -
\\
&&-2a^{4}\frac{Q^{2}}{\left( r_{0}\right) ^{4}}\cos ^{2}\theta \left\{ \frac{%
17}{11}\left( -1\right) ^{1/4}\frac{r}{r_{0}}F\left[ iArc\text{Sinh}\left[
\left( -1\right) ^{1/4}\frac{r}{r_{0}}\right] ,-1\right] \right\}
\end{eqnarray*}
where the constants have been selected to obtain assymptotically the
Kerr-Newman metric and $P_{ST}$ is identical to the similar quantity in the
static case:
\[
P_{ST}=\frac{1}{3}Q^{2}\left\{ \overline{r}^{4}+\overline{r}^{2}\sqrt{%
\overline{r}^{4}+1}+2\overline{r}\left( -1\right) ^{1/4}F\left[ Arc\text{sin}%
\left[ \left( -1\right) ^{3/4}\overline{r}\right] ,-1\right] \right\} \,\ \
\ ;\ \ \ \ \ \left( \ \overline{r}\equiv \frac{r}{r_{0}}\right)
\]

We can see that this solution contains new terms that do not appear in the
Kerr-Newman model as in Reissner-N\"{o}rdstrom and the static Born-Infeld
model. There are products of charge and angular momentum. The expansion is
for $a/r<<1$\ and $r_{0}\lesssim r$\ . Is useful to remark here that the $%
a_{KN}$ appear as a parameter into the constant of integration $B\left(
\theta \right) $ (in the same manner that $-2M$ into the constant $A\left(
\theta \right) $). The parameter $a$ is the parameter of {\it deformation of
the spherical} {\it symmetry }in the Boyer and Lindquist type interval, eq.
(1) .

\section{Analysis of the metric in the Born-infeld rotating case}

The general behaviour of the metric is similar to the Kerr-Newman's model
(almost globally). As one can see from the last expression for the $\Delta $%
, the metric has two horizons and depends strongly on $r_{0}$ (related to
Born-Infeld parameter $b^{2}\equiv Q^{2}/\left( r_{0}\right) ^{4}$) and its
quotient with $a^{2}$. The assymptotical behaviour of the $\Delta \left(
r\right) $ is:
\[
\Delta \left( r\right) \cong r^{2}-\left[ 2M+\frac{Q^{2}}{r_{0}}\left( \frac{%
4}{3}1.854-1.525\frac{2a^{2}}{r_{0}^{2}}-\frac{2a^{4}}{r_{0}^{4}}%
2.8653\right) \right] r+Q^{2}+a_{KN}^{2}+a^{2}\left( 0.8576\frac{a^{2}Q^{2}}{%
r_{0}^{4}}\right)
\]
that corresponds to have asymptotically an ADM mass:
\[
M_{ADM}=M+\left[ \frac{Q^{2}}{r_{0}}\left( \frac{2}{3}1.854-1.525\frac{a^{2}%
}{r_{0}^{2}}-\frac{a^{4}}{r_{0}^{4}}2.8653\right) \right]
\]
and an ADM angular momentum:
\[
a_{ADM}=\sqrt{a_{KN}^{2}+0.8576\frac{2a^{4}Q^{2}}{r_{0}^{4}}}
\]
where $M$ and $a_{KN}$ are the two constants of integration that appear in
the Kerr-Newman model (these constants are related to the asymptotic values
of mass and angular momentum in the Maxwellian-linear case of a black hole
[8]). Notice that the numerical coefficient 1.854 is characteristic of the
static EBI monopole and arises from the leading terms in the expansion of
the hypergeometric function $F$. The numerical coefficient 1.525 is part of
the integration constant $A\left( \theta \right) $ and is obtained from the
asymptotic conditions imposed on the rotating EBI model.

\section{Conclusions}

In this report a solution of the Einstein-Born-Infeld equations for slowly
rotating black holes is presented. The general behaviour of the geometry is
strongly modified according to the value that takes $r_{0}$ (Born-Infeld
radius[1,2]) relative to $a$ value. This metric permits solutions of $%
M^{2}<a_{ADM}^{2}+Q^{2}$ , $M=0$ and $\ a_{KN}=0$ with a regular horizon
(see Figures 1, 2 and 3). The spacetime of the Born-Infeld-rotating monopole
(in contrast to Maxwell counterpart) have intrinsic or particular ADM values
of mass and angular momentum. In this non-linear electromagnetic rotating
model, the mass and angular momentum of the spacetime($a_{ADM}$ and $M_{ADM}$%
) can be driven by: the electromagnetic charge $Q$, the absolute field of
Born-Infeld $b$ (or the BI radius: $r_{0}$) and the $a$ parameter from the
geometry. Solutions with $M<0$ exist, strongly bounded for the $M_{ADM}$
value, this $M_{ADM}$ value cannot to be negative because naked
singularities appear in the spacetime, violating in this manner the positive
mass theorem for Riemannian (assymptotically flat) manifolds [10,11].
In order to clarify some points related with the electromagnetic fields
and the topology of the manifolds, we can make the following comments:
the problem in the integrability conditions for the metric, that carry us to make
a "slowly rotating" expansion (i.e.
 $a/r<<1$ ), there exist also in the
Bianchi identitity for the electromagnetic fields. For the limit in which the
metric solution is integrable and remains valid(i.e. $a/r<<1$), the
Bianchi identity also remains valid, obviously because it is precisely the
geometrical caracter of the electromagnetic field tensor (closed form),
and depends on the geometry of the space-time where these
electromagnetic fields are living. All the remarks and computations
are absolutely consistent with the "slowly rotating regime".
From the point of view of the Hamilton-Jacobi equations and the Petrov
classification for this type of non-linear rotating solutions, a good
analisys of this problem was made in [16].The origin of this difficulties concerning on the integrability and a
cure for this problem will be given in a forthcoming work where we will
analyze the exact rotating solution[17].

\section{Acknowledgements}

I am very grateful to all the people of the Bogoliubov Laboratory of
Theoretical Physics and Directorate of JINR for their hospitality and
support. I am very thankful to Professor Yurii Stepanovsky, Georgii Afanasiev
and Boris Barbashov for their guide in my scientific formation and very
useful discussions. I appreciate deeply Professor Daniel Sudarsky's efforts
to clarify to me many concepts in on the mass as geometrical propierty of
the manifolds.

\section{References}

[1] B.Hoffmann, Phys. Rev.{\bf 47}, 887 (1935).

[2] M.Born and L. Infeld, Proc. Roy. Soc. (London), {\bf 144}, 425 (1934).

[3] M.Born, Proc. Roy. Soc. (London), {\bf 143}, 411 (1934).

[4] L.D. Landau and E.M. Lifshitz, {\it Teor\'{\i}a cl\'{a}sica de los Campos%
}, (Revert\'{e}, Buenos Aires, 1974), p. 504.

[5] C. Misner, K. Thorne and J.A. Weeler, {\it Gravitation}, (Freeman,
SanFrancisco,1973), p. 474.

[6] D. Harari and C. Loust\'{o}, Phys. Rev.D{\bf \ 42}, 2626 (1990).

[7] A. Borde,`` Regular Black Holes and Topology Change'', gr-qc/9612057.

[8] S. Chandrasekhar, {\it The Mathematical Theory of Black Holes}, (Oxford
University Press, Oxford, 1992, p. 632.

[9] D.J. Cirilo Lombardo, {\it The axially symmetric geometry with
Born-Infeld fields: Rotating Einstein-Born-Infeld model}, Tesis de la
Licenciatura en Ciencias F\'{\i}sicas (undergraduate tesis), Universidad de
Buenos aires, September of 2001, 61 pp.

[10] R. Shoen and S. T. Yau, Comm. Math. Phys. {\bf 65}, 1457 (1979).

[11] R. Penrose, Ann. New York Acad. Sci{\bf .224}, 125 (1973).

[12] D.J. Cirilo Lombardo, Problems of Atomic Science and Technology 6, 71
(2001).

[13] R. Metsaev and A. Tseytlin, Nucl. Phys. B 293, 385 (1987).

[14] O. Andreev and A. Tseytlin, Nucl. Phys. B 311, 205 (1988).

[15] D.J. Cirilo Lombardo, Preprint JINR-E2-2003-221.

[16] D.J. Cirilo Lombardo, The Newman–Janis algorithm, rotating
solutions and Einstein–Born–Infeld black holes, Class. Quantum
Grav. 21 (2004) 1407–1417 (and references therein).

[17] D.J. Cirilo-Lombardo (in preparation).

\end{document}